\documentclass[sigconf]{acmart}
\usepackage{caption}
\usepackage{subcaption}
\usepackage{float} 
\usepackage[ruled,vlined,linesnumbered]{algorithm2e}
\usepackage[most]{tcolorbox}
\usepackage{graphicx}
\usepackage{xargs} 
\usepackage{subfiles}
\usepackage{paralist}
\usepackage{comment}
\usepackage{todonotes}
\usepackage[noabbrev]{cleveref}
\acmConference[Arxiv]{}{2025}{Salt Lake City, USA}

\renewcommand\footnotetextcopyrightpermission[1]{} 
\begin{document}
\title{cuTeSpMM: Accelerating Sparse-Dense Matrix Multiplication using GPU Tensor Cores}
\definecolor{keywordcolor}{rgb}{0.5,0,0.1}
\author{Lizhi Xiang}
\affiliation{%
 \institution{University of Utah}
 \streetaddress{}
 \city{Salt Lake City}
 \state{Utah}
 \country{United States}
 \postcode{}
}
\email{u0814474@utah.edu}

\author{Omid Asudeh}
\orcid{0000-0002-1825-0097}
\affiliation{%
 \institution{University of Utah}
 \streetaddress{}
 \city{Salt Lake City}
 \state{Utah}
 \country{United States}
 \postcode{}
}
\email{asudeh@cs.utah.edu}

\author{Gerald Sabin}
\orcid{0000-0001-5109-3700}
\affiliation{%
 \institution{RNET Technologies}
 \city{Dayton}
 \state{Ohio}
 \country{United States}
}
\email{gsabin@rnet.com}

\author{Aravind Sukumaran-Rajam}
\affiliation{%
 \institution{Meta Platforms}
 \city{}
 \state{California}
 \country{United States}
}
\email{aravindsr@meta.com}

\author{P. Sadayappan}
\affiliation{%
 \institution{University of Utah}
 \streetaddress{}
 \city{Salt Lake City}
 \state{Utah}
 \country{United States}
 \postcode{}
}
\email{saday@cs.utah.edu}

\begin{abstract}
Many recent GPUs feature matrix multiplication engines (aka Tensor Core Units or TCUs) that perform small fixed-size matrix-matrix products at very high throughput. They have been used very effectively to speed up dense matrix-matrix multiplication libraries like Nvidia's cuBLAS, enabling significantly higher performance over use of the traditional scalar GPU cores. There also been recent interest in using these dense TCUs for the important sparse-dense matrix-matrix multiplication (SpMM) kernel via explicit zero-filling.
However, an examination of the attainable performance of TC-GNN, the state-of-the-art TCU-enhanced SpMM  implementation, indicates that for a substantial majority of the sparse matrices in the SuiteSparse collection, the achieved performance falls significantly short of the state-of-the-art SpMM kernels that only utilize scalar cores.

In this paper, we therefore address the question: {\bf Can dense TCUs be effectively used to accelerate SpMM for a range of sparse matrices arising from multiple application domains, such as those found in the SuiteSparse matrix collection?} We answer this question in the affirmative by developing a very efficient TCU-based GPU kernel - cuTeSpMM (cuda Tensor core SpMM) that achieves substantially higher performance over TC-GNN. We also develop a notion of the {\em TCU-Synergy} of a sparse-matrix, based on its non-zero structure and a modeled Operational Intensity. For sparse matrices with high TCU-synergy, cuTeSpMM outperforms state-of-the-art scalar-core SpMM implementations, while achieving only slightly lower performance on matrices with low TCU-Synergy.

\end{abstract}
\maketitle

\section{Introduction}
\label{sec:intro}


\begin{figure}[!ht]
\centering
\begin{minipage}{.25\textwidth}
\includegraphics[width=0.9\linewidth]{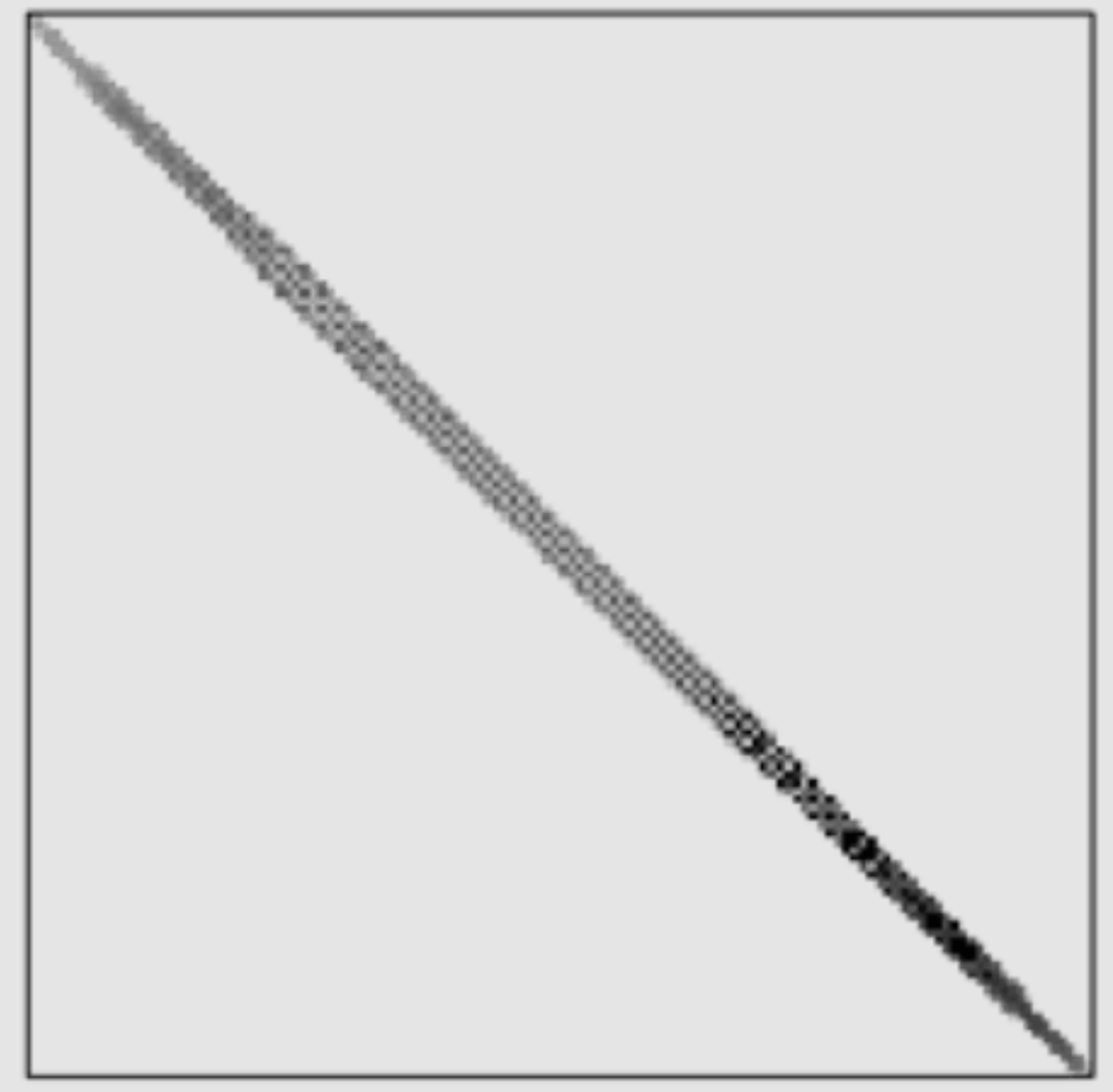}
\caption*{$Emilia\_923$}
\label{fig:emilia_923}
\end{minipage}%
\begin{minipage}{.25\textwidth}
\includegraphics[width=0.9\linewidth]{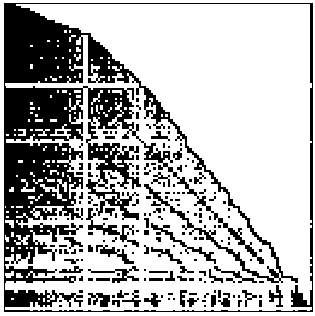}
\caption*{$NotreDame\_www$}
\label{fig:notreDame_www}
\end{minipage}
\caption{Nonzero patterns of two sparse matrices from the
SuiteSparse matrix collection \cite{davis2011university, suitesparse}}
\label{fig:2sparseMats}
\end{figure}

Matrix engines are now being incorporated in several recent CPUs (e.g., IBM Power 10 \cite{IBM}, Intel Sapphire Rapids \cite{Intel}) and GPUs (e.g., Nvidia Ampere \cite{a100TC} and Hopper \cite{hopperTC}, AMD MI250 and MI300\cite{AMD}). These matrix engines enable very fast multiplication of small fixed-size matrices through new instruction set extensions and direct hardware implementation for fixed-size matrix multiplication. 
These matrix engines, also called Tensor Cores, have been used to significantly enhance the performance of vendor libraries like cuBLAS \cite{nvidia2024cublas} that implement dense linear algebra operations.

{\em Can these high throughput, fixed size, dense tensor core hardware implementations be used for sparse-dense matrix multiplication (SpMM)?} While sparse matrices used in scientific computing and data analytics tend to be extremely sparse (with far less than 1\% of elements being
non-zero), the distribution of non-zeros is generally not uniformly random across the 2D index space. The non-zero distribution pattern tends to exhibit clustering in sub-regions of the 2D index space, which leaves open the possibility of developing zero filling strategies to leverage these dense TC Units. Fig.~\ref{fig:2sparseMats} shows the non-zero pattern for two sparse matrices from the SuiteSparse matrix collection \cite{davis2011university,suitesparse}. The matrix $Emilia\_923$ arose from a structural application and has $923136$ rows/columns and $41005206$ nonzero elements (i.e., it is extremely sparse with a density of less than $0.005\%$). However the non-zero elements tend to be clustered around the diagonal. If the 2D $923136 \times 923136$ index space is partitioned into a regular grid of $16 \times 4$ ``bricks'' (this is one of the supported dense matrix sizes in Nvidia's WMMA API for using its tensor cores for matrix-matrix multiplication), the average number of non-zero elements in non-empty bricks is nearly 20\% dense.
$NotreDame\_www$
 is a web graph with $325760$ rows/columns and $929849$ nonzero elements, with an extremely low density of less than $0.0009\%$. Its non-zeros are scattered over a much wider range of the 2D index space, but even for this much sparser and more scattered matrix, the average density of non-zero elements in non-empty $16 \times 4$ bricks is nearly 10\%. Therefore, there is the potential to develop algorithms to leverage dense tensor core hardware to accelerate sparse matrix computations. 

 The examples above illustrate that although sparse matrices arising in applications from scientific computing and data analytics may be extremely sparse, the non-zero patterns can be concentrated in subregions of the 2D index space. As explained later, further compaction into ``bricks'' corresponding to the matrix operand sizes of TCU instructions can further increase the density of non-zeros in non-empty bricks. Thus, there has been interest in developing SpMM implementations that utilize TCUs \cite{ref:blkELL,wang2023tc,zachariadis2020accelerating}. TC-GNN \cite{wang2023tc} is a recently developed TCU-based implementation of Graph Neural Networks (GNNs), where the dominant computational operation is SpMM. Wang et al. \cite{wang2023tc} demonstrated their TCU-based SpMM implementation (which we refer to simply as TC-GNN in this paper instead of TC-GNN-SpMM) to be superior to other TCU-based SpMM implementations. Since SpMM comparisons with other scalar-core (SC) based GPU implementation of SpMM were not performed by Wang et al. \cite{wang2023tc}, we carried out experimentation on around 1000 matrices (all matrices with more than 10,000 rows) from the SuiteSparse collection \cite{suitesparse,davis2011university}. We compare the performance of TC-GNN with the fastest scalar-core implementation (best among cuSparse \cite{cusparse}, GeSpMM \cite{huang2020ge}, Sputnik \cite{gale2020sparse}) on this wide range of sparse matrices. 

\begin{figure}[!ht]
\centering

  \begin{subfigure}[b]{0.49\columnwidth} 
    \includegraphics[width=\linewidth]{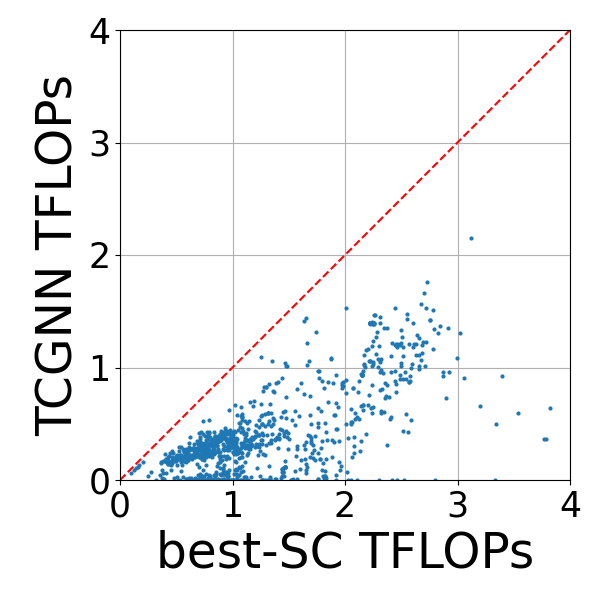}
  \end{subfigure}
  \hfill 
  \begin{subfigure}[b]{0.49\columnwidth} 
    \includegraphics[width=\linewidth]{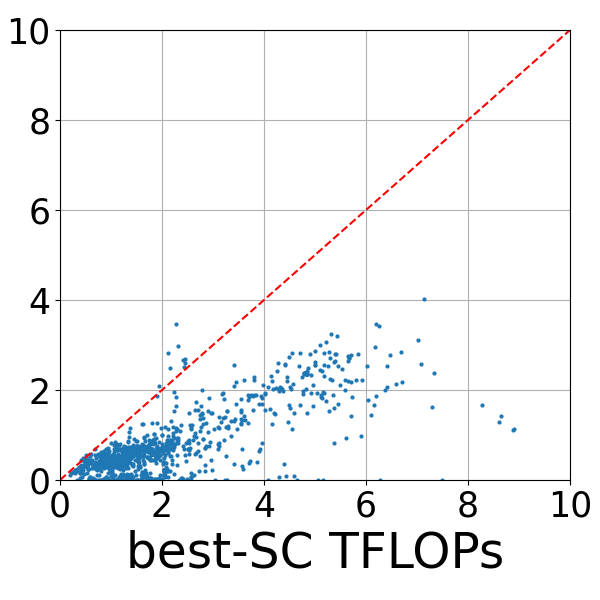}
  \end{subfigure}


\caption{TC-GNN performance versus the best SC performance (cuSparse \cite{cusparse}, GeSpMM \cite{huang2020ge}, Sputnik \cite{gale2020sparse}). Left: Ampere a100, right:RTX-4090}
\label{fig:tcgcc-bestSC}
\end{figure}

 Fig.~\ref{fig:tcgcc-bestSC} contains two scatter-plots depicting performance in TFLOPs achieved by TC-GNN vs. the best scalar-core SpMM (labeled best-SC). The top chart shows performance on an Nvidia RTX 4090 GPU, while the bottom chart shows performance on an NVidia Ampere A100 GPU. The width of the dense matrices was 128. The peak scalar-core 32-bit floating-point performance (FP32) on the A100 (4090) is 19.2 TF (82.6 TF), while the TCU peak performance (TF32) is 156 TF (A100) and 82.6 TF (4090). On the 4090 GPU, the performance of TC-GNN is worse than best-SC except for a handful of matrices. On the surface, this is not entirely surprising as the peak TCU throughput is equal to the peak SC throughput on the RTX 4090. Therefore, the handful of TCU ``wins" must arise from other architectural and algorithm design differences. 
 In contrast, the A100 has an 8x higher peak TCU throughput as compared to the A100 peak scalar-core throughput. Despite the significantly higher peak TCU performance of the A100, the performance of TC-GNN is relatively worse when compared to best-SC; TC-GNN was not faster on any of the 1000+ matrices! For these matrices, TC-GNN is not able to exploit the 8x higher peak throughput of the TCU.

Given these experimental results on the state-of-the-art TC-GNN SpMM, we seek to answer the following two questions in this paper:
\begin{enumerate}
    \item Is it possible to design and implement a TCU-based GPU SpMM that is actually faster than the best scalar-core SpMM on a non-trivial fraction of SuiteSparse matrices?
    \item Can some inherent characteristics of a sparse matrix be identified that indicate whether it has good potential for a faster TCU-based SpMM implementation than the best-SC SpMM?
\end{enumerate}

 In this paper, we successfully address both these questions.  We develop a new TCU-based implementation of SpMM - cuTeSpMM (cuda Tensor core SpMM) that achieves substantially higher performance over the state-of-the-art TC-GNN TCU-based SpMM kernel that out performs the best SC SpMM implementations for a large number of matrices. We also develop a notion of the TCU-Synergy of a
sparse-matrix. The TCU-Synergy is based on a matrix's non-zero structure, and it characterizes a given sparse matrix with respect to its potential for acceleration of SpMM via TCU versus scalar core implementations. 

This paper makes the following contributions:
\begin{itemize}
    \item We devise a new sparse matrix representation - Hierarchical Row-Panel-Blocking (HRPB) - designed to enable high performance SpMM on Nvidia GPUs with dense tensor cores.
    \item We develop a high-performance SpMM kernel for the HRPB sparse matrix representation called cuTeSpMM.
    \item Using 1000+ sparse matrices from the SuiteSparse matrix collection, we demonstrate significant performance improvements over the state-of-the-art TC-GNN \cite{wang2023tc} SpMM.
    \item For the first time (to our knowledge), we demonstrate the benefits of using Nvidia GPU tensor cores for achieving much higher SpMM performance on a significant fraction of SuiteSparse matrices than that achievable by the best scalar-core SpMM.
    \item We develop a TCU-Synergy metric, model the Operational Intensity of cuTeSpMM based on the Synergy, and demonstrate the correlation between Synergy, modeled OI, and SpMM throughput.
\end{itemize}
\section{Background}

\label{sec:background}

Before we present details on the design of the key data structure and algorithms of cuTeSpMM, we provide background information on dense tensor cores and how they can be used for the sparse SpMM computation.


NVIDIA's Tensor Cores are hardware units designed to accelerate deep learning computations. 
A  GPU Tensor Core unit performs an $m$ x $k$ x $n$ Matrix Multiply-Add (MMA) operation $D$ = $A$ x $B$ + $C$, where $A$ is of size $m \times k$, $B$ is of size $k \times n$, and $C$ and $D$ are of size $m \times n$. The size of $m$, $k$, and $n$ and the supported data type of each vary based on the specific GPU architecture.  
In this paper, we focus on SpMM operations for scientific computing and data analytics applications, where 32-bit precision is commonly used.
$TF32$ is a new floating point format, introduced with the Tensor Core units, 
with the same range as the IEEE standard 32-bit floating-point (8 bits for representing the exponent) but lower precision (10 bits for the mantissa). In order to use the tensor core units, the dot product computation (which forms the building block for both matrix multiplication and convolutions), rounds $FP32$ inputs to $TF32$, computes the product of two $TF32$ inputs into an $FP32$ register, and then accumulates those products into an $FP32$ output\cite{tf32, tf32:1}. This approach reduces the input data size to 21 bits, while preserving the output precision of $FP32$. 

Tensor core matrix multiply-accumulate (MMA) operations are warp-synchronous, meaning all threads within a warp (i.e., a group of 32 threads) collectively participate in the operation. Each thread is responsible for computing a portion of the output matrix $D$ using rectangular slices of matrices $A$ and $B$~\cite{andrewkerr}. The supported sizes of the matrix subsets are architecture dependent. For example, the Ampere $TF32$ tensor core MMA supports an $A$ shape of ${(16, 4)}$, a $B$ shape of ${(4, 8)}$, and an output matrix ($D$) shape of ${(16, 8)}$. 
During tensor core MMA operations, every consecutive groups of 4 threads form a thread group within a warp, resulting in 8 thread groups. Each thread within a group loads an element from a row of the A operand matrix. Given the A operand matrix's 16 rows, each thread group cyclically loads two rows. For the B matrix operand, each thread group loads a column from the B matrix into its registers. Regarding the C operand matrix, each thread group stores two rows of the resulting matrix. Within each row of C, every thread in a group holds two elements. 

Developers can use tensor cores via the high level $cublas.gemmex$ library API or by using the lower level $nvcuda::wmma::*$  namespace API
from the $mma.h$ cuda header file. 
To implement a basic Matrix Multiply one can divide the output between the warps such that each warp is responsible for generating a small tile of the output matrix. For each output tile the relevant input tiles are distributed amongst the threads of the warp in which each thread is owning a fraction.

\section{cuTeSpMM Design}
\label{sec:design}

\begin{figure*}[htbp]
  \centering
  \begin{subfigure}[b]{0.58\textwidth}
    \includegraphics[width=\linewidth]{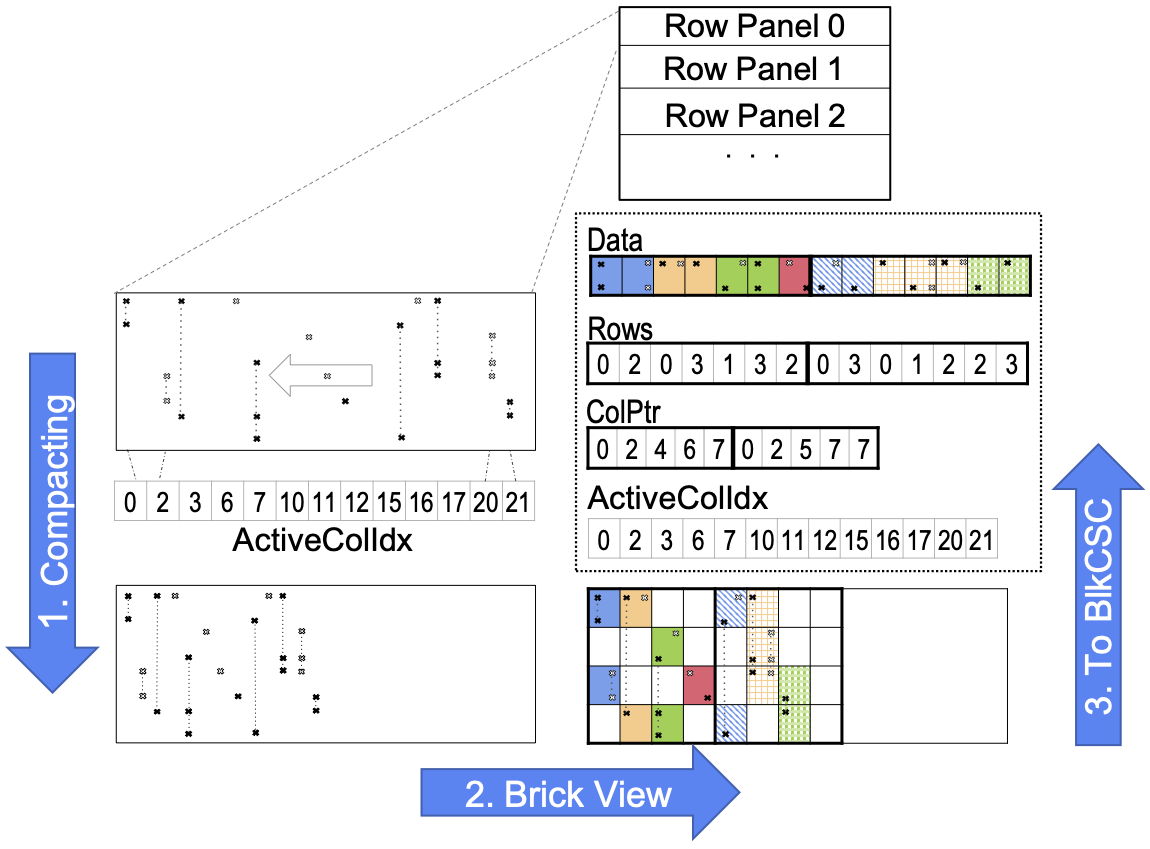}
    \caption{ }
    \label{fig:testa}
  \end{subfigure}
  \hspace{20pt}
  \begin{subfigure}[b]{0.28\textwidth}
    \includegraphics[width=\linewidth]{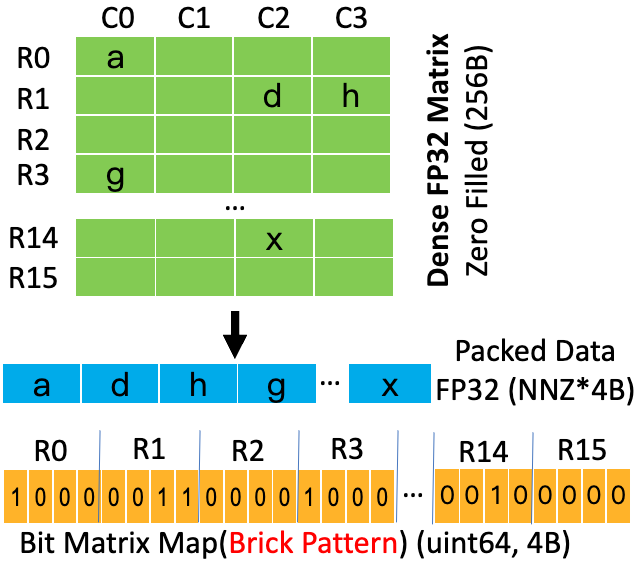}
    \caption{ }
    \label{fig:testb}
  \end{subfigure}
  \caption{Sparse Compression Process}
  \label{fig:data-struct-csc}
\end{figure*}

\lstdefinestyle{mystyle}{
    language=C,
    basicstyle=\small\ttfamily, 
    captionpos=b,
    keywordstyle=\color{keywordcolor},
    numbers=left,
    morekeywords={__syncthreads, shared, Write, unrolled, FMAs},
    numbersep=5pt, 
}
\begin{figure}[ht!]
\lstset{
    style=mystyle,
}
\begin{lstlisting}
struct Block{
    float* nnz_array;
    uint64 patterns [TM/brick_m][TK/brick_k];
    uint colPtr[TK/brick_k + 1];
    uint rows[TM/brick_m * TK/brick_k];
}
\end{lstlisting}
\caption{Block Data Strucutre}
\label{fig:block data structure}
\end{figure}

\begin{figure}[ht!]
\lstset{
    style=mystyle,
}
\begin{lstlisting}
struct HRPB{
    Byte * packedBlocks;
    uint blockedRowPtr [M/TM + 1];
    uint activeCols[NUM_BLKS * TK];
    uint sizePtr[NUM_BLKS + 1];
}
\end{lstlisting}
\caption{HRPB data structure}
\label{fig:HRPB}
\end{figure}



\subsection{Notations}
\label{sec:notations}


In this paper, we use the following notation: $A$ denotes the sparse input matrix, $B$ denotes the dense input matrix, and $C$ denotes the dense output matrix. $M$ and $K$ represent the number of rows and columns of matrix $A$, respectively. $N$ represents the width of matrix $B$. We refer to a contiguous set of rows of $A$ as a row-panel, and we denote the size of each row panel as $TM$. Each row panel is partitioned along the column dimension into blocks of shape $(TM, TK)$, where $TK$ is the tile size along the column dimension. Each block is further sub-divided into bricks of shape $(brick\_m, brick\_k)$. The matrix $B$ is divided into bricks of shape $(brick\_k, brick\_n)$. Each tensor core operation computes the product of $A[brick\_m][brick\_k]$ with $B[brick\_k][brick\_n]$ and accumulates the results to the output brick $C[brick\_m][brick\_n]$. 

\subsection{HRPB Sparse Matrix Data structure}
\label{sec:cuTe_data}


The sparse representation significantly influences performance in multiple aspects. An effective sparse representation must minimize overall storage requirements, thereby reducing DRAM usage and, crucially, data volume. Additionally, the data structure should facilitate efficient decompression to generate the dense representation necessary for tensor cores. Given that the total number of tensor core operations correlates directly with the total number of tiles containing at least one non-zero value, an efficient sparse representation should aim to decrease the number of such tiles. Our HRPB (Hierarchical Row-Panel-Blocking) data structure design is guided by these principles. \Cref{fig:data-struct-csc} shows a high-level overview of our design. We divide the rows into row panels. Within each row panel, we aggregate all columns containing at least one non-zero value and organize them into blocks. Each such block is then compressed to store only the non-zero values. The rest of the section details this process. 

The first step in creating the HRPB is to  collect (re-order) all active columns in a row panel and place them together towards the left. A column in a row panel is considered active if it contains at least one non-zero element. These columns are then grouped to form bricks. This process is illustrated in the ``compacting`` step of Figure \ref{fig:data-struct-csc} (a). Note that this step reduces the number of active blocks and thus the total operations. Each row panel is divided into blocks of shape $(TM, TK)$, where we set $TM$ either 16 or 32. The $TK$ is set to 16 based on our experiments. On Nvidia Ampere architecture, for TF32 data type, the tensor core operations support fragments (slices) of shape (16, 4) for the A matrix, shape (4, 8) for the B matrix, and the output shape is (16, 8) of the C matrix. Hence, we further subdivide the blocks into bricks of shape ${(brick_m, brick_k)}$, where $brick_m$ is set to 16, $brick_k$ is set to 4, and $brick_n$ is set to 8. The compaction step helps to reduce the number of active blocks and bricks, which in turn minimizes the number of FLOPS. To maintain the column index information after compacting, we maintain an array called ${ActiveColIdx[]}$ for each row panel. The $i$-th element of this array represents the original column id of the $i$-th active column.

The active bricks in a block are stored in a column-major format, which closely resembles the Compressed Sparse Column (CSC) format. The ``To BlkCSC'' component of Figure \ref{fig:data-struct-csc} (a) illustrates our data structure. For each active brick, we encode the non-zero distribution pattern, i.e., the indices corresponding to non-zero elements, into a 64-bit vector. Since the brick size is 64, every bit in the 64-bit vector indicates the corresponding position of the brick is non-zero or not. For instance, pattern[i] indicates if the ith element in the brick is non-zero or not. Figure \ref{fig:data-struct-csc} (b) illustrates this process. These patterns are stored in an array called `patterns'. Additionally, we utilize an array named ${nnz\_array}$ to maintain the non-zero values across different active bricks in CSC order. The non-zero values within each brick are ordered in row major and appended to the ${nnz\_array}$ of block. In addition to the ${nnz\_array}$ and ${patterns}$, we utilize a 'rows' array to indicate the row ID of each active brick and a 'colPtr' array to denote the starting index of each brick column. Figure \ref{fig:block data structure} illustrates the structure of a block. The 'colPtr', 'patterns', and 'rows' arrays are collectively referred to as metadata.


Figure \ref{fig:HRPB} illustrates the HRPB data structure, which organizes all the blocks in the sparse matrix. The `packedBlocks' in the HRPB data structure represent a memory chunk containing all the blocks of all the row panels packed together. Different row panels may contain  different number of blocks. To identify all blocks belonging to individual row panels, we utilize an array called `blockedRowPtr'. This array has a size of ${\frac{M}{TM} + 1}$ and resembles the row pointer array in the CSR format, storing the starting block index of each row panel. For instance, blockedRowPtr[i] indicates the starting block index of row panel i, and blockedRowPtr[i+1] - blockedRowPtr[i] indicates the number of blocks inside row panel `i'. However, since each block may contain a varying number of non-zero elements, their memory requirement may differ. Hence, `blockedRowPtr' alone cannot precisely locate the position of a block within the `packedBlocks' memory chunk. Hence, we use sizePtr$[NUM\_BLKS + 1]$ to determine the starting memory address of each block in `packedBlocks'.
The `activeCols' array stores the active column indices of each row panel. 

\begin{algorithm}[t]
\LinesNumbered
\caption{\textsf{cuTeSpMM} kernel design}
\label{alg:kernel}
\small
\textbf{Input:} {blockedRowPtr[], sizePtr[], packedBlocks[], numRowPanels}\\
\textbf{Output:} {matrix C[]}\\
${\textbf{shared}}$  ${SM\_A[TM*TK], SM\_B[TK][N]}$\\
${a\_frag} [2]$\\
${b\_frag} [4]$\\
${c\_frag} [4][TM/brick_m][4]$\\
${warp\_id = threadIdx/32}$\\
${lane\_id = threadIdx\%32}$\\
${group\_id = lane\_id/4}$\\
${tid\_in\_group = lane\_id\%4}$\\
${row\_panel = threadblock\_id}$\\
${block\_start = blockedRowPtr[row\_panel]}$\\
${block\_end = blockedRowPtr[row\_panel + 1]}$\\
\For{$block\_id = block\_start$ \KwTo $block\_end$}{
\text{$block\_addr = sizePtr[block\_id]$}\\
\text{$block\_addr\_end = sizePtr[block\_id + 1]$}\\
\text{$SM\_A = packedBlocks[block\_addr:block\_addr\_end]$}\\
\text{$block = (Block)SM\_A$}\\
\For{$i = warp\_id$ \KwTo TK \KwSty{step} $num\_warps$}{
\text{$b\_row = block.active\_cols[i]$}\\
 \For{$j = lane\_id$ \KwTo N \KwSty{step} 32}{
    \text{$SM\_B[i][j] = B[b\_row * N + j]$}
    }
}
\text{syncthreads}\\
\text{$nnz\_offset = 0$}\\
\For{$i = 0$ \KwTo TK/ $brick_k$}{
    \For{$n = 0$ \KwTo 4}{
            \text{$b\_frag[n] = SM\_B[i*brick_k + tid\_group]$}\\
            \text{$[warp\_id*4*brick_n+n*brick_n+group\_id]$}
        }
    \For{$j = block.colPtr[i]$ \KwTo $block.colPtr[i+1]$}{
         \text{$afrag[0] = afrag[1] = 0 $}\\
        \text{$row = block.rows[j]$}\\
        \text{$pattern = block.patterns[j]$}\\
        \If{pattern[lane\_id]}{
            index = count\_1s[pattern[0:lane\_id]]\\
            afrag[0] = block.nnzs[nnz\_offset + index]
        }
        \If{pattern[lane\_id + 32]}{
            index = count\_1s[pattern[0:lane\_id + 32]]\\
            afrag[1] = block.nnzs[nnz\_offset + index]
        }
        \text{$nnz\_offset += count\_1s(pattern)$}\\
        \For{$n = 0$ \KwTo 4}{
            \text{$warp\_wmma(a\_frag, b\_frag[n],$} \\ 
            \text{$c\_frag[n][row])$}\\
        }
    }
}
\text{syncthreads}
}
\text{$m = threadblock\_id * TM$}\\
\text{n = $4 * warp\_id * brick_n$}\\
\text{Write(C[m: m+ TM][n:n+$brick_n*4$],$c\_frag$)}\\
\end{algorithm}

\subsection{cuTeSpMM Kernel Design}
\label{sec:cuTe kernel }

\begin{figure}[ht!]
\centering
\includegraphics[width=0.8\linewidth]{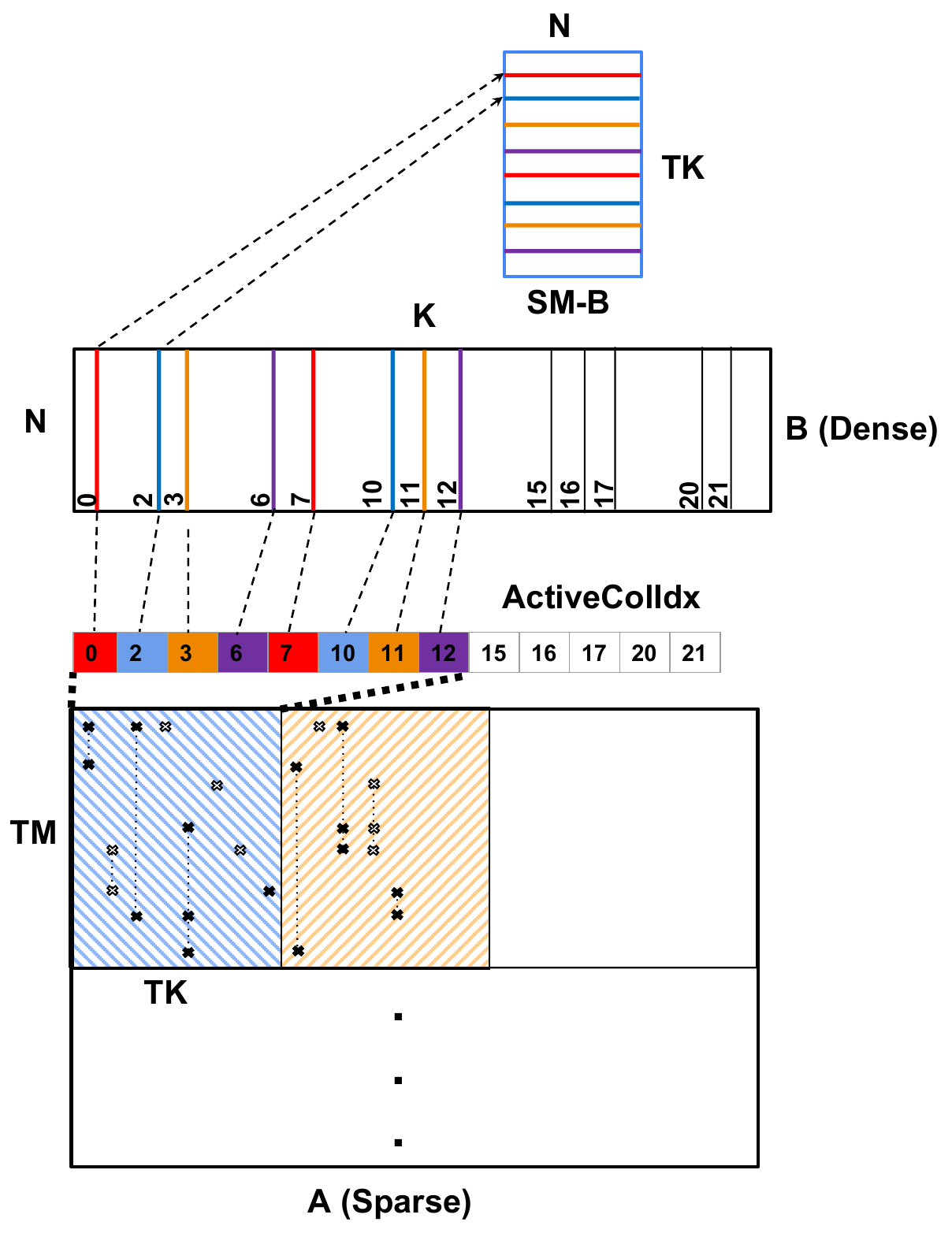}
\caption{Data Transfer from Dense Matrix into the Shared Memory}
\label{fig:smem-trans}
\end{figure}


The performance of GPUs is influenced by various parameters, including data movement, load balance, and the number of useful computations. As previously stated, our brick-compressed representation aids in reducing the total number of FLOPS. This section discusses how our cuTeSpMM kernel decreases data movement by leveraging the GPU memory hierarchy without compromising occupancy. We will also elaborate on how we achieve effective load balance and minimize thread idling. Algorithm \ref{alg:kernel} outlines our overall scheme.

A thread block is designated to handle all non-zero blocks within each row panel. For simplicity, let's consider a feature size of 128, where each thread block comprises 128 threads (4 warps). The precise approach for determining the number of threads and thread blocks per row panel will be explained in detail later. 
%
The different warps in a thread block are distributed along the feature dimension (N). Each warp produces a ${TM \times 32}$ tile of C. The number 32 is also referred to as TN in the subsequent text. The four warps collectively compute a C tile of size ${TM \times 128}$. Note that each tensor core operation only produces a $16 \times 8$ output tile of C, but each warp computes a ${TM \times 32}$ output tile of C. This warp coarsening has two main benefits: firstly, it enhances the register-level reuse of the A matrix, and secondly, it reduces the number of brick decode operations, which involve multiple non-zero bit scanning operations. However, excessive warp coarsening can decrease occupancy and parallelism. We determined the coarsening factor as 4 through empirical experiments. In this configuration, the kernel grid size is (${\frac{M}{TM}, \frac{N}{128}}$).


For small feature sizes, like 32, allocating a thread block of size 128 to process each block would lead to inefficient resource utilization. Therefore, for smaller feature sizes, we decrease the number of warps in a thread block to ${\frac{N}{warp\_coarsening\_level \times 8}}$ (where the 8 represents the tensor core MMA operand tile size along N, referred to as $brick\_n$). For large N size, a row panel is assigned multiple thread blocks, while each thread block produce a ${TM \times 128}$ portion of the output matrix. For simplicity, for the later discussion, we will assume that  N is 128, which means the thread block size is 128.


Algorithm \ref{alg:kernel} illustrates the kernel design. Each thread sequentially iterates over the blocks along the reduction dimension K inside the row panel (Line 14). For each block, we initially load the non-zeros of the active bricks inside the block and the corresponding meta-data into shared memory $SM_A$ in a coalesced manner with the assistance of sizePtr (line 17). Next, we unpack A in line 18. Given that the maximum non-zero inside the block is ${TM \times TK}$, within the GPU thread block, we allocate ${TM \times TK + MetaDataSize}$ shared memory space to store the block. Following that, we load the required rows of B into shared memory. We utilize the ${block.active\_cols}$ array with a size of TK to identify the necessary rows of B. We fetch the corresponding rows from global memory in a coalesced manner and pack them into a contiguous 2D buffer in shared memory (Lines 19-22). Fig. \ref{fig:smem-trans} illustrates the process of loading the necessary rows of B into the shared memory buffer.


After loading the necessary rows of B to shared memory, we will traverse the active bricks inside the block in a CSC manner. Inside a block, there will be ${\frac{TK}{brick_k}}$brick columns. For each brick column, the block.colPtr indicates the starting and ending active brick index of this brick column. The  block.rows pointer is used to figure out the brick row index and the brick.patterns is to find out the brick non-zero pattern by a given active brick index. As we explained earlier, the non-zero brick pattern is a 64-bit vector. 
In line 34, each thread in a warp calculates the number of non-zero elements assigned to the threads with a lane id less than the current thread. This index represents the position of the non-zero element that this thread should load in the current brick.
%
%
%
Within the block, multiple bricks will exist. To keep track of the starting position of non-zeros in each active brick, we introduce the ${nnz\_offset}$ variable. Initially set to 0, ${nnz\_offset}$ is incremented after decoding a brick pattern by the number of non-zeros in the brick. Thus, ${nnz\_offset}$ indicates the starting position of non-zeros for the next brick to be processed. The scanning of prefix bits is highly efficient and is implemented using bitwise operations and is computed by the CUDA cores, which do not participate in the core MMA computations (Tensor Cores handle these computations).

After reading the corresponding non-zeros into registers, a warp collectively processes the four distinct output tiles of the C matrix within the loop (line 29). Inside the loop, each thread in the warp is required to load the corresponding element from the B matrix in shared memory, denoted as $SM_B$, into registers. Following this, the thread performs the MMA operation and stores the partial results in ${cfrag}$. Once all the blocks in a row panel have been iterated, each thread will write its ${cfrag}$ to global memory. This implies that the thread block will collectively write out a ${TM \times 128}$ tile of the C matrix to global memory.

\section{Analysis and Parameter Selection}

\label{sec:analysis}

As discussed in Section \ref{sec:design}, cuTeSpMM loads $A$ (the sparse matrix) and $B$ (the dense matrix) from shared memory into global memory, and then from shared memory into registers. This allows for efficient coalesced global memory accesses, and explicit guaranteed reuse from shared memory. The dense output ($C$) is held stationary in registers and written directly into global memory, and therefore its memory usage is implicitly minimized by design. 

We denote the density of the packed HRPB bricks as ($\alpha$), where $\alpha=nnz_b/(brick_m*brick_k)$. Since the HRPB compression removes empty columns in each brick, the $alpha$ is both the average brick density as well as the average density of a column in a brick. Further, the number of bricks ($B$) is $ nnz/(\alpha*brick_m*brick_n)$.

In modern Nvidia GPU architectures, each memory transfer loads 128 bytes (32 4-byte words) of contiguous data. HRPB arranges data to optimize loads of the sparse matrix $A$. As described in Section \ref{sec:design}, each brick includes a 64-bit ``non-zero pattern" and the 4-byte non-zero values. 
Each thread must load the 8 byte mask (2 transactions) and the collective warp must read the non-zeros, therefore the number of shared memory to register transactions for a brick is $shared\_mem\_trans_a=\lceil (\alpha*brick_m*brick_k)/32 \rceil+2$. Therefore, the total shared memory reads for the entire $A$ matrix is  $[(((\alpha*brick_m*brick_k)/32) \rceil+2)*\frac{N}{TN}] * nnz/(\alpha*brick_m*brick_k)$. 
The total number of shared memory transactions for A is:

\begin{equation}
    shmem\_trans_A
     \approx \frac{N*nnz*4}{\alpha*TN*brick_m*brick_k}
    \label{eq:a_loads}
\end{equation}

One row of the dense $B$ matrix is loaded from shared memory into registers once for each packed column in HRPB
A row has N elements, and requires $(N*4)/128$ shared-memory load instructions. Therefore, the total number of shared memory loads into registers is $(N/8)*brick\_k*NUM\_ACTIVE\_BRICKS(nnz/(\alpha*brick_m*brick_k))$, see Equation \ref{eq:b_loads}.
\begin{equation}
 shmem\_trans_B=\frac{N*nnz}{32*\alpha*brick_m}
 \label{eq:b_loads}
\end{equation}

Considering the specific case of 16 for $brick_m$ and 4 for $brick_k$ required for the WMMA operation on our target GPUS (Nvidia A100 and RTX 4090), we get:
\begin{equation}
shmem\_trans_A \approx \frac{N*nnz}{\alpha*16*TN}; shmem\_trans_B=\frac{N*nnz}{512*\alpha}
\end{equation}
Increasing $TN$ decreases the number of shared-memory loads for $A$. However, the register usage per thread increases proportionately with $TN$ and the total number of thread-blocks decreases proportionately as $TN$ is increased. We choose a value of $TN$ that makes the total number of shared-memory transactions for $A$ and $B$ equal, i.e., $TN$=32.

For the choice of $TN$=32, the operational intensity (ratio of the total number of compute operations over the number of load operations) is:
\begin{equation}
OI_{shmem}=\frac{2*N*nnz}{\frac{N*nnz}{512*\alpha}+\frac{N*nnz}{512*\alpha}}=512*\alpha
\label{eq:oi_1}
\end{equation}

\label{sec:TM_TK}


\begin{figure}[htbp] 
  
  \begin{subfigure}[b]{0.49\columnwidth} 
    \includegraphics[width=\linewidth]{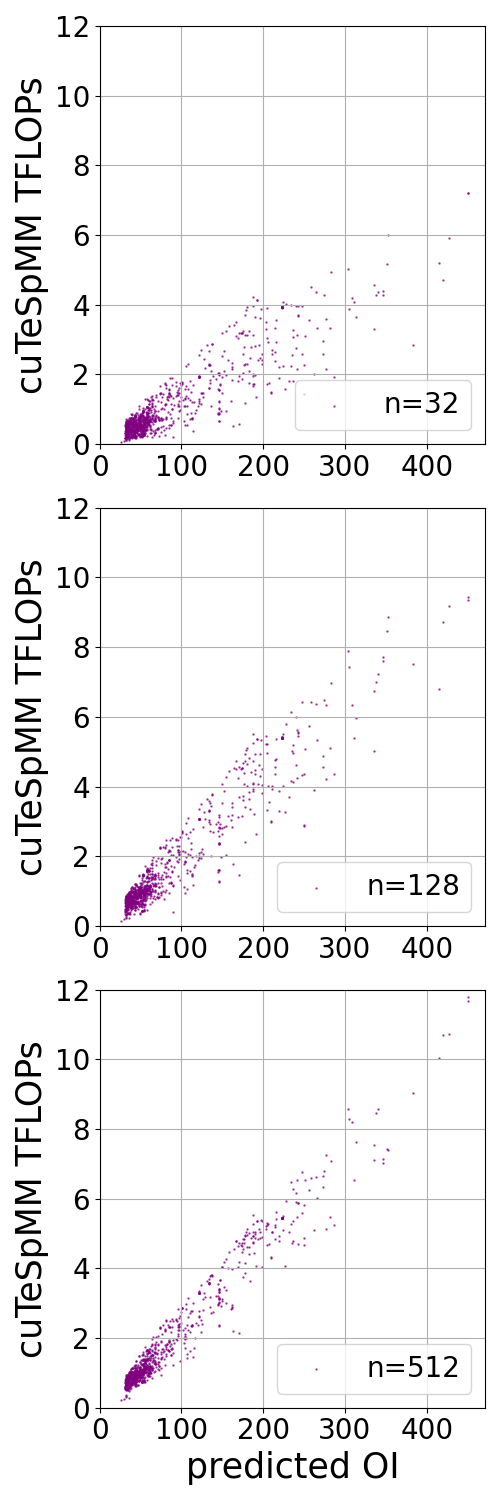}
  \end{subfigure}
  \hfill 
  \begin{subfigure}[b]{0.49\columnwidth} 
    \includegraphics[width=\linewidth]{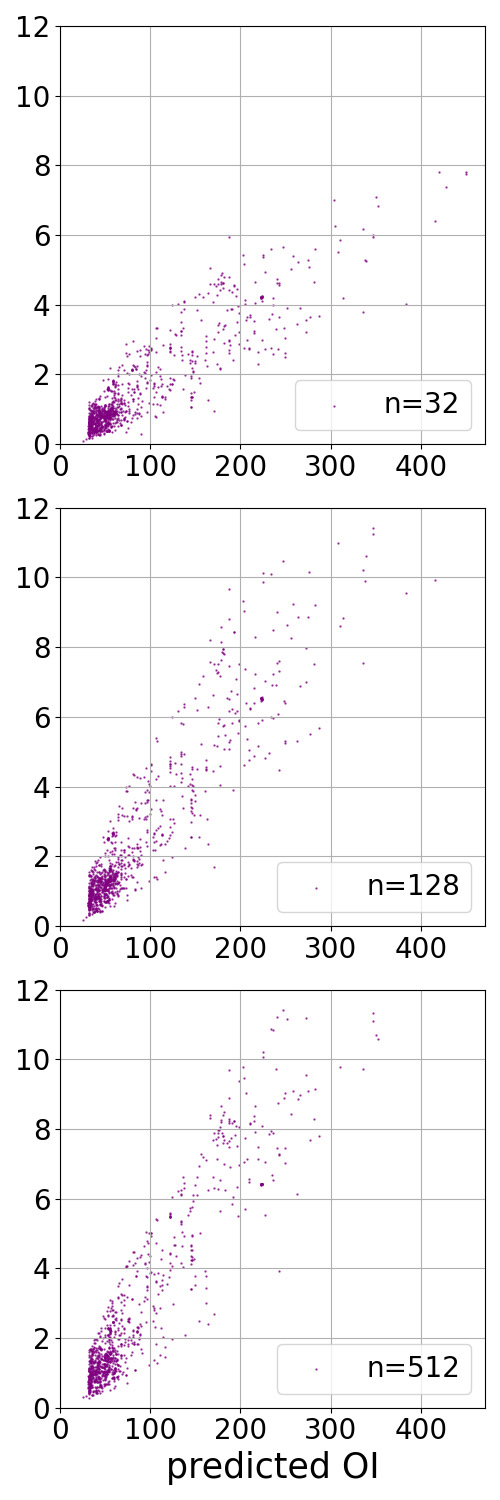}
  \end{subfigure}
    \caption{Modeled $OI_{shmem}$ is well correlated to GFLOPs for cuTeSpMM across a large number of matrices and multiple dense matrix widths (Left: Ampere A100, right: RTX-4090). 
    }
    \label{fig:modeled_oi}
\end{figure}

While detailed experimental evaluation is presented later, we show experimental data in Fig. \ref{fig:modeled_oi}, for achieved cuTeSpMM performance on the 1000+ tested matrices on the A100 and RTX 4090 GPUs, to demonstrate that the modeled $OI_{shmem}$ is quite strongly correlated with measured performance. The figure plots  $OI_{shmem}$ ($512*\alpha$) versus the measured performance (in TFLOPs) for $N$ = 32, 128 and 512. We believe that the HRPB data-structure and the effective choice of $TN$ to optimize $OI_{shared}$ are key reasons for the high performance achieve dby cuTeSpMM.

\begin{figure}[ht!]
\centering
\includegraphics[width=0.85
\linewidth]{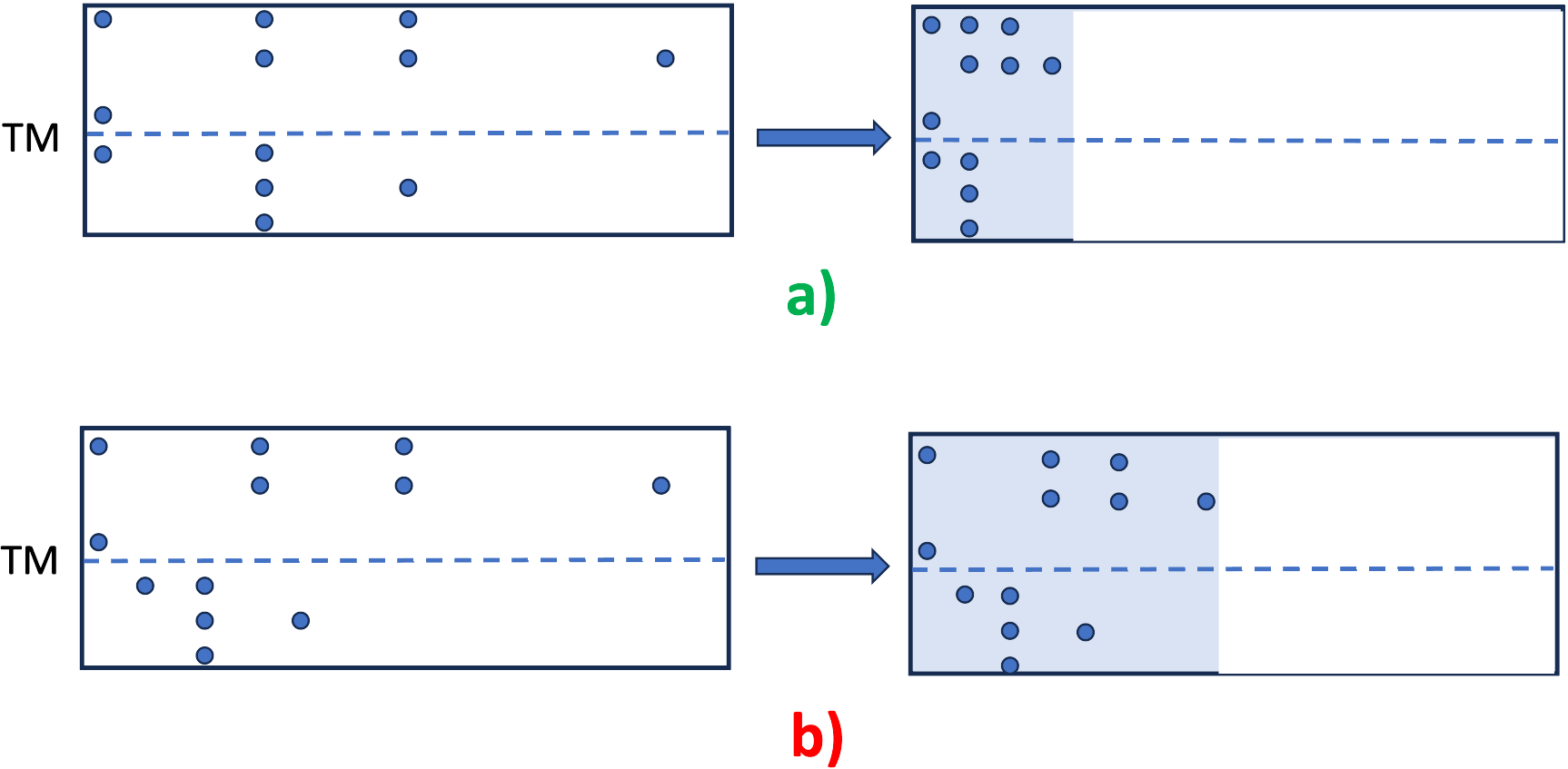}
\caption{NNZ-distribution after left packing}
\label{fig:tm_packing}
\end{figure}

The cuTeSpMM algorithm has parameters, (${TM}$, ${TK}$) representing the tile sizes along the dimensions $M$ and $K$ of blocks within a row panel. The $TK$ size will not have an affect on $\alpha$, thus, it doesn't directly impact data movement and the number of FMA computations. However, it does influence other performance factors. In line 25 of our kernel design algorithm, we traverse all ${\frac{T_K}{brick_k}}$ brick columns inside a block. This for loop is fully unrolled since both ${TK}$ and ${brick_k}$ are constants. The prefix scanning operations used to decode the non-zero distribution pattern are entirely independent. A larger ${TK}$ value enhances ILP (Instruction-Level Parallelism), potentially improving performance. However, a larger ${TK}$ value also increases the volume of shared memory required, and thus reduces occupancy(line 3 in \cref{alg:kernel}).Increased shared memory usage can reduce occupancy, potentially degrading parallelism. Quantifying ILP is challenging as the compiler affects the generated code. We empirically determined that a ${TK}$ size of 16 balances the occupancy and ILP.
In the previous ${OI}$ analysis, we assumed that ${TM}$ equals ${brick_m}$, signifying that every time we load a brick of the sparse A matrix, we must load the corresponding elements of B from shared memory to register in order to perform tensor core HMMA. If ${TM}$ is larger than ${brick_m}$, each brick column may contain ${\frac{TM}{brick_m}}$ bricks. The active bricks in a brick column require the same elements from the B matrix. Consequently, the B elements loaded from shared memory to register can be reused across all active bricks within a brick column. Let's denote $\beta$ as the average number of bricks inside a brick column of a block. The number of shared memory transfers of the B matrix can then be reduced to:
\begin{equation}
shmem\_trans_B=\frac{Nnnz}{32 *\alpha *brick_m *\beta}
\label{eq:b_loads_tm}
\end{equation}
assuming the brick column density $\alpha$ remains constant. Thus, increasing $T_M$ directly increases the $OI_{shmem}$ due to increased reuse. However, this increase in reuse is often counteracted by a decrease in the column non-zero density $\alpha$ as $T_M$ increases. For a sparse matrix, a large ${TM}$ often leads to a drop in brick column density.
\Cref{fig:tm_packing} illustrates two row panels, each with a height of ${TM}$ but featuring different non-zero distributions. The dashed line divides a row panel into two sub-row panels, each with a height of ${\frac{TM}{2}}$. In the first row panel, as depicted in \Cref{fig:tm_packing} (a), the active columns of the two sub-row panels are common. Left active column packing in this scenario results in a densely packed region highlighted in light blue. However, in the second row panel (\Cref{fig:tm_packing} (b)), the active columns of the two sub-row panels are disjoint. Consequently, after left packing, the dense region for the second row panel is much sparser than that of the first row panel. Furthermore, since the active columns of the two sub-row panels are entirely disjoint, there is no data reuse across the two sub-row panels. Hence, for the second row panel, it is more beneficial to set the row panel size to ${\frac{TM}{2}}$ and assign the two sub-row panels to different thread blocks. These two row panels represent the two extreme cases of different non-zero distributions within a row panel. In reality, the non-zero distribution of a row panel typically falls somewhere between these two extreme cases.
 Although a large ${TM}$ value has the potential to increase the data reuse of ${B}$ elements, thereby increasing $OI$, it also comes with a disadvantage. This disadvantage arises from the fact that a large ${TM}$ increases shared memory and register usage, which degrades occupancy. Consider that the matrices from suitesparse are super sparse, and the non-zero distribution is far away from the distribution shown in \Cref{fig:tm_packing} a, and the disadvantage of the occupancy drop for large $TM$. In our evaluation, all the matrices we evaluated are using $TM$ = 16, which is the size of ${brick_m}$.


\section{Load balancing improvement}
\label{sec:balancing}

In the prior discussion on kernel design, we explored the assignment of thread blocks to row panels. One challenge we need to address is load balance. Certain row panels may contain numerous active columns, while others may only feature a few. Simply assigning thread blocks directly to row panels in these situations can lead to a considerable load imbalance. Thread blocks assigned to lighter row panels may finish their tasks much earlier than those assigned to heavier row panels, resulting in the inefficient utilization of GPU resources.


One effective strategy to address load imbalance is to manage the scheduling of thread blocks. This involves sorting the row panels in descending order according to their workload (number of active columns) and scheduling the work in a way that prioritizes the heaviest panels for execution first. Although the initial thread blocks assigned to heavier panels may necessitate more processing time and occupy specific streaming multiprocessors (SMs) for an extended duration, the remaining SMs can concurrently handle lighter row panels efficiently. This approach enables SMs to continue processing lighter tasks while the first few SMs focus on their heavier counterparts, thereby optimizing GPU resource utilization. However, this approach has a significant disadvantage: A lot of sparse matrices display a structure in which consecutive row panels feature similar active columns. In such instances, assigning consecutive thread blocks to process consecutive row panels can enhance the reuse of the B matrix and potentially boost cache hits. Rearranging the order of row panels might disrupt this structure and lead to an increase in cache misses.



An alternative approach is to enhance load balance by dividing the workload of heavy row panels. A row panel is deemed heavy if the associated workload surpasses the average. These heavy row panels are divided along the K dimension, creating multiple virtual row panels. The number of blocks in these virtual row panels is adjusted to be equal to the average number of blocks across all row panels. Nevertheless, this strategy introduces another challenge. When the row panel is divided into multiple virtual row panels along the K dimension, we must employ atomic operations to consolidate partial results from these various virtual panels. 

Our load-balancing scheme is a variation of the latter approach, taking into account both the number of GPU waves and the cost of atomic operations. Each SM can concurrently handle a specific number of thread blocks. If the total number of thread blocks exceeds ${NUM\_SMs \times NUM\_THREAD\_BLK\_PER\_SM}$, these thread blocks will be distributed across multiple waves. The number of waves can be determined as \newline ${\frac{Total\_thread\_blocks}{NUM\_SMs \times NUM\_THREAD\_BLK\_PER\_SM}}$. This calculation relies on the kernel's resource requirements, such as shared memory and register usage. To illustrate the significance of waves, consider the following example. Suppose we have 991 row panels, requiring 991 thread blocks if we don't partition the row panels. Additionally, let's assume there are 100 SMs, and each SM can accommodate one thread block. Therefore, we would have ceil(991/100) = 10 GPU waves. Now, let's consider a scenario where the first row panel is heavy, consisting of 10 blocks, while the others contain only 1 block each. In total, we have 1000 blocks. The SM responsible for executing the first-row panel has 10 blocks to process, while the remaining 99 SMs each have 990 row panels to process, with 1 block in each panel. Consequently, all SMs evenly process 10 blocks. This example illustrates that even when dealing with a heavily loaded row panel, workload distribution across the SMs can remain balanced without the need to partition the heavy row panel.

Inspired by this example, we design our load balancing scheme as follows: First, for each row panel, we compute the $num\_loads$ of that row panel. The $num\_loads$ is defined as the equation:
\begin{equation}
num\_loads = \frac{num\_blocks\_of\_row\_panel}{AVG\_BLK\_ROW\_PANEL}.
\label{eq:num_loads}
\end{equation}

Next, we compute the partition ratio of each row panel by equation \ref{eq:partition_ratio}. 

\begin{equation}
partition\_ratio = \frac{num\_loads}{num\_waves}.
\label{eq:partition_ratio}
\end{equation}

Then, for each row panel, we partition it into multiple row panels based on the $partition\_ratio$ (no partition happens if $partition\_ratio = 1$). 

To compute the $num\_waves$ in equation \ref{eq:partition_ratio}, we need to determine the number of thread blocks that can be resident in an SM concurrently and the number of SMs. This information can be obtained during the compilation phase by querying the device (GPU) information. We also need the ${Total\_thread\_blocks}$ which is determined at runtime. 




Compared with the second approach, which partitions the row panel only based on $average\_blocks\_each\_row\_panel$, our strategy reduces the partition ratio by the factor of $num\_waves$, translating to a reduction in the number of atomic operations by the factor of $num\_waves$.




\section{Experimental Evaluation}
\label{sec:experiments}

\subsection{Hardware, Datasets, and Compared Algorithms}
\label{sec:setting}

\begin{figure*}[htbp]
  \centering
  \noindent\begin{minipage}{\textwidth} 
    \centering
    \includegraphics[width=0.5\textwidth]{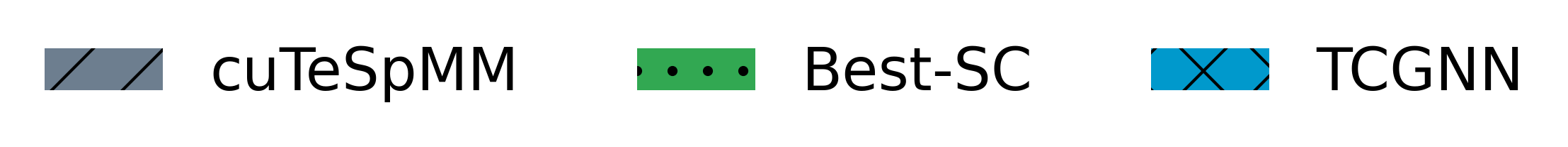}
  \end{minipage}
  
  \begin{subfigure}[b]{0.32\textwidth}
    \includegraphics[width=\linewidth]{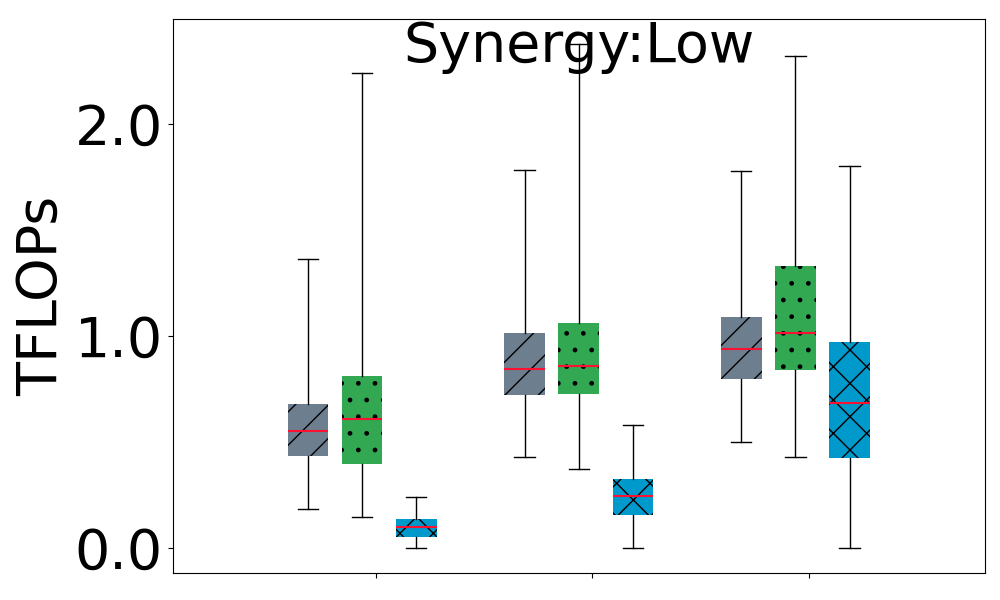}
  \end{subfigure}
  \begin{subfigure}[b]{0.32\textwidth}
    \includegraphics[width=\linewidth]{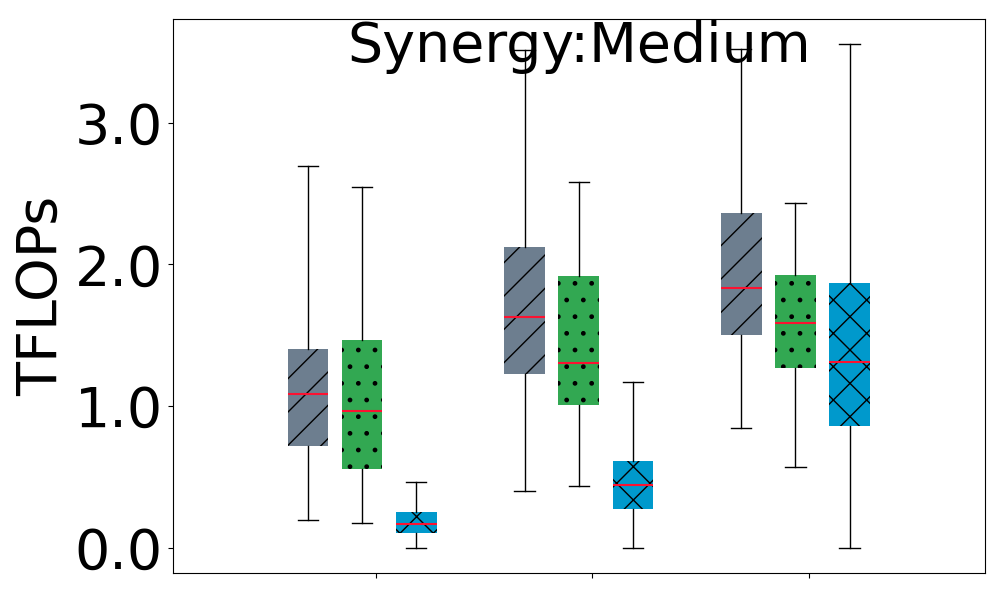}
  \end{subfigure}
  \begin{subfigure}[b]{0.32\textwidth}
    \includegraphics[width=\linewidth]{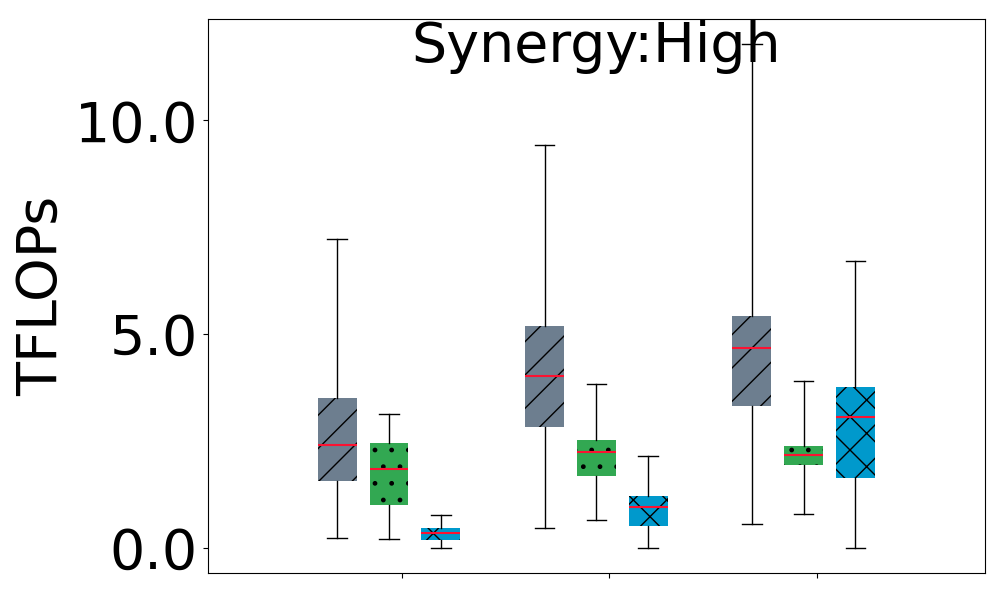}
  \end{subfigure}
  
  \begin{subfigure}[b]{0.32\textwidth}
    \includegraphics[width=\linewidth]{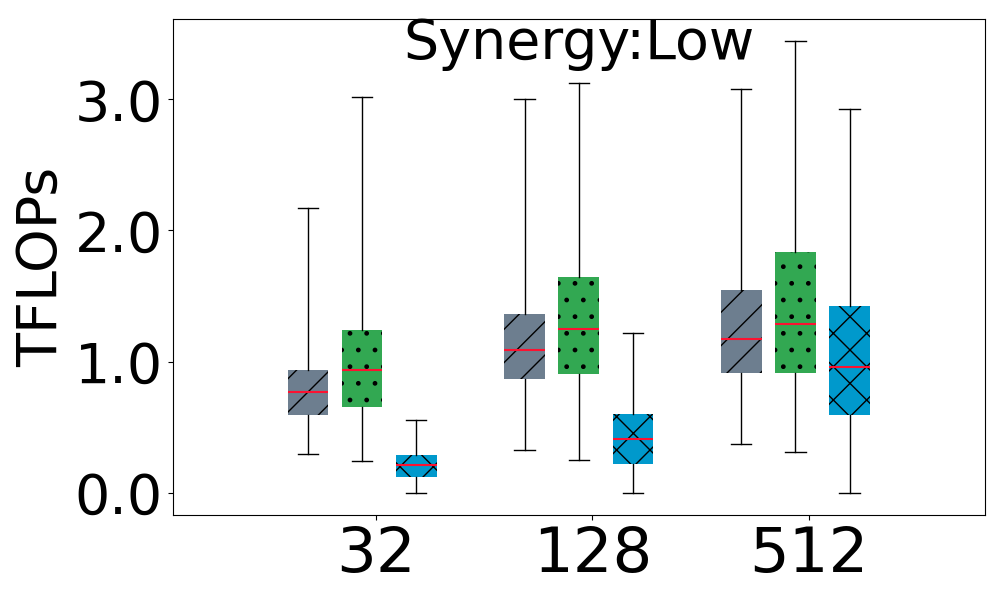}
  \end{subfigure}
  \begin{subfigure}[b]{0.32\textwidth}
    \includegraphics[width=\linewidth]{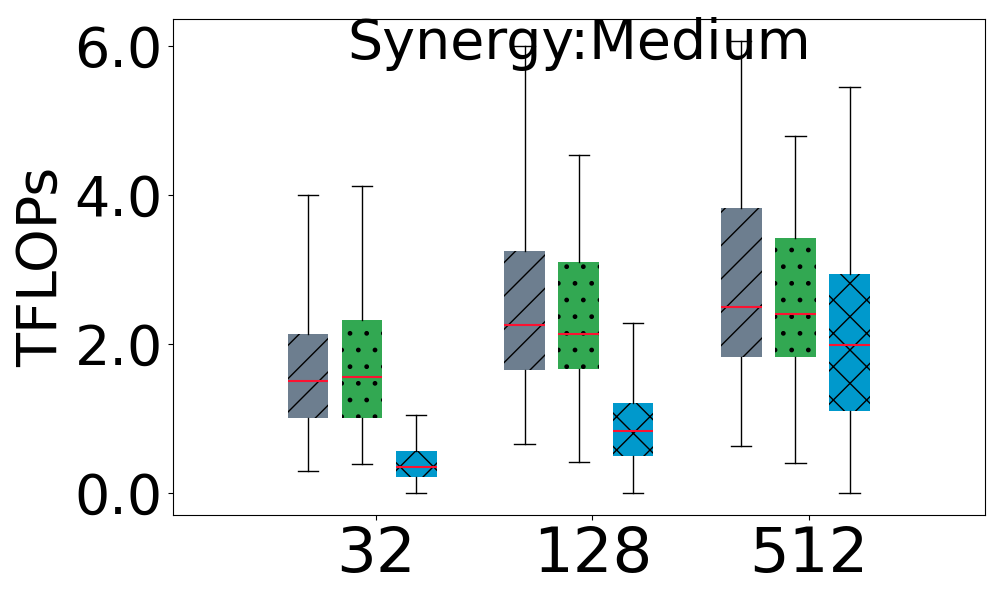}
  \end{subfigure}
  \begin{subfigure}[b]{0.32\textwidth}
    \includegraphics[width=\linewidth]{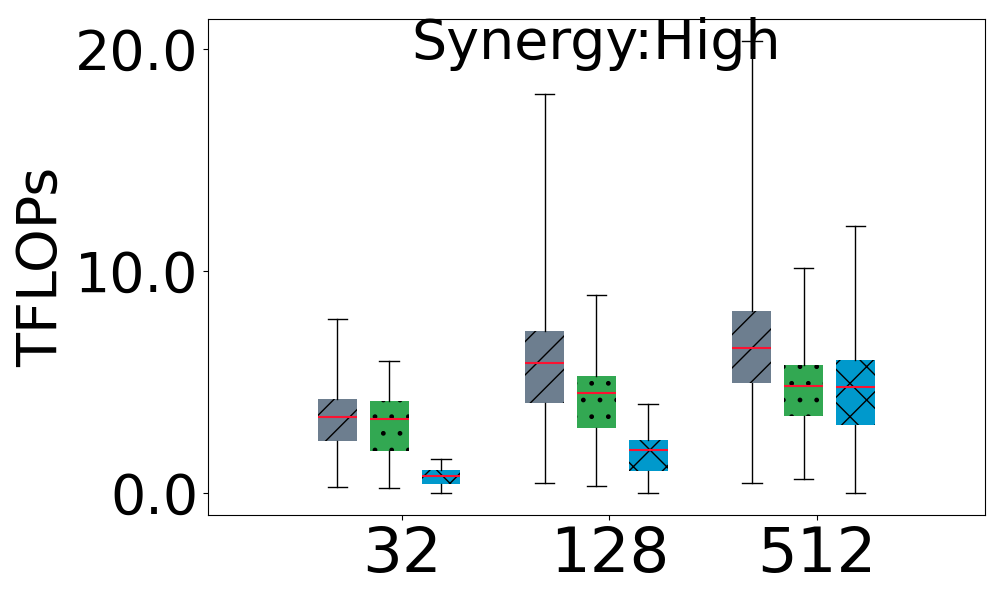}
  \end{subfigure}
  
  \caption{Comparing cuteSpMM, best-CUDA, and TCGNN on different synergy groups for Ampere a100 (upper row) and RTX-4090 (lower row) on three dense matrix size width (32, 128, 512)}
  \label{fig:combined_boxplots}
\end{figure*}

\begin{figure}[htbp] 
  \centering
  \includegraphics[width=1\columnwidth]{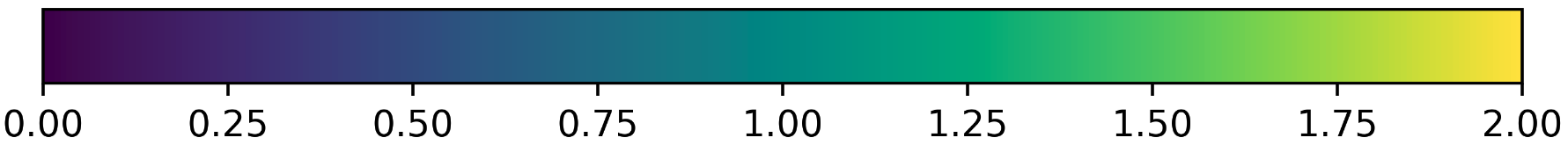}
  
  \begin{subfigure}[b]{0.49\columnwidth} 
    \includegraphics[width=\linewidth]{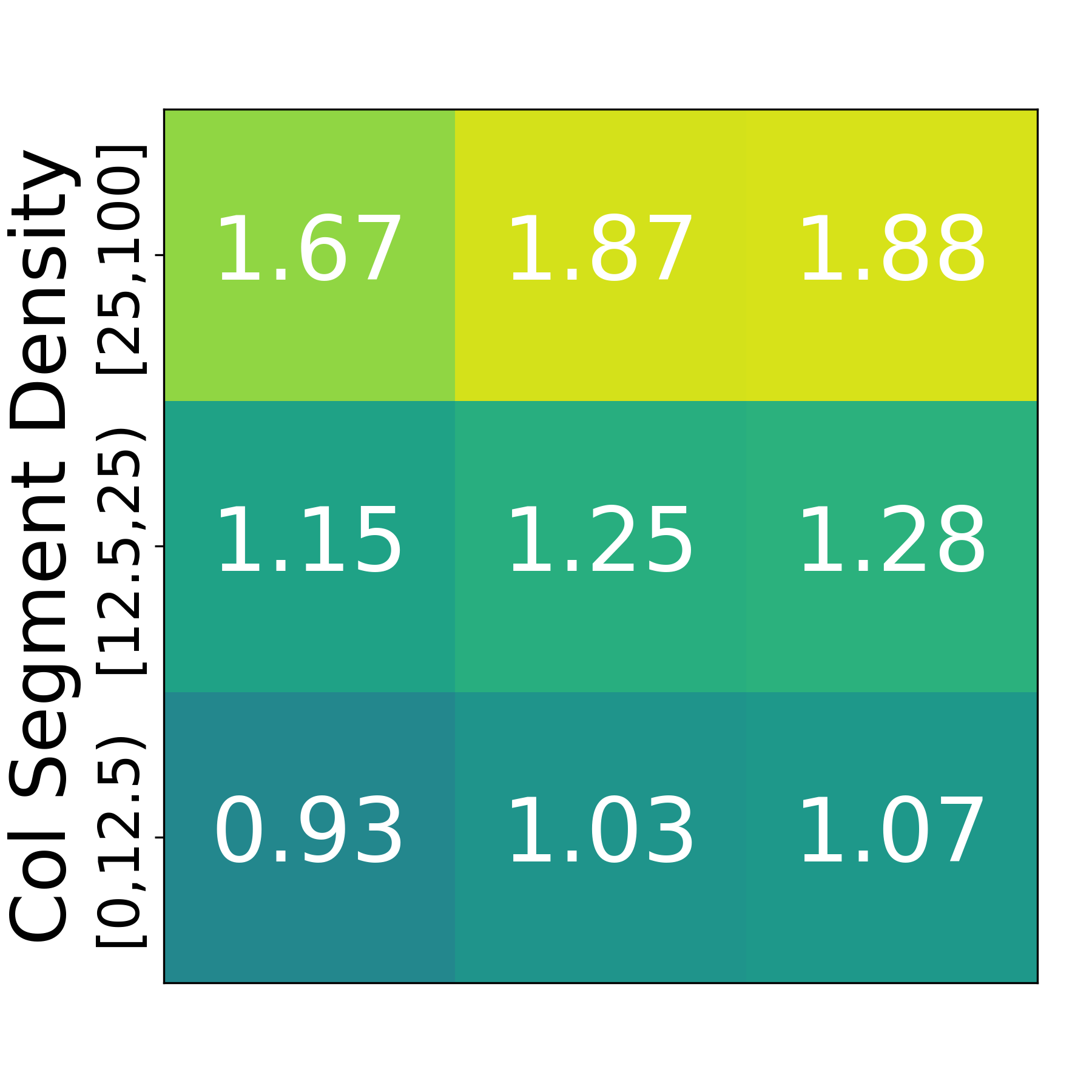}
  \end{subfigure}
  \hfill 
  \begin{subfigure}[b]{0.49\columnwidth} 
    \includegraphics[width=\linewidth]{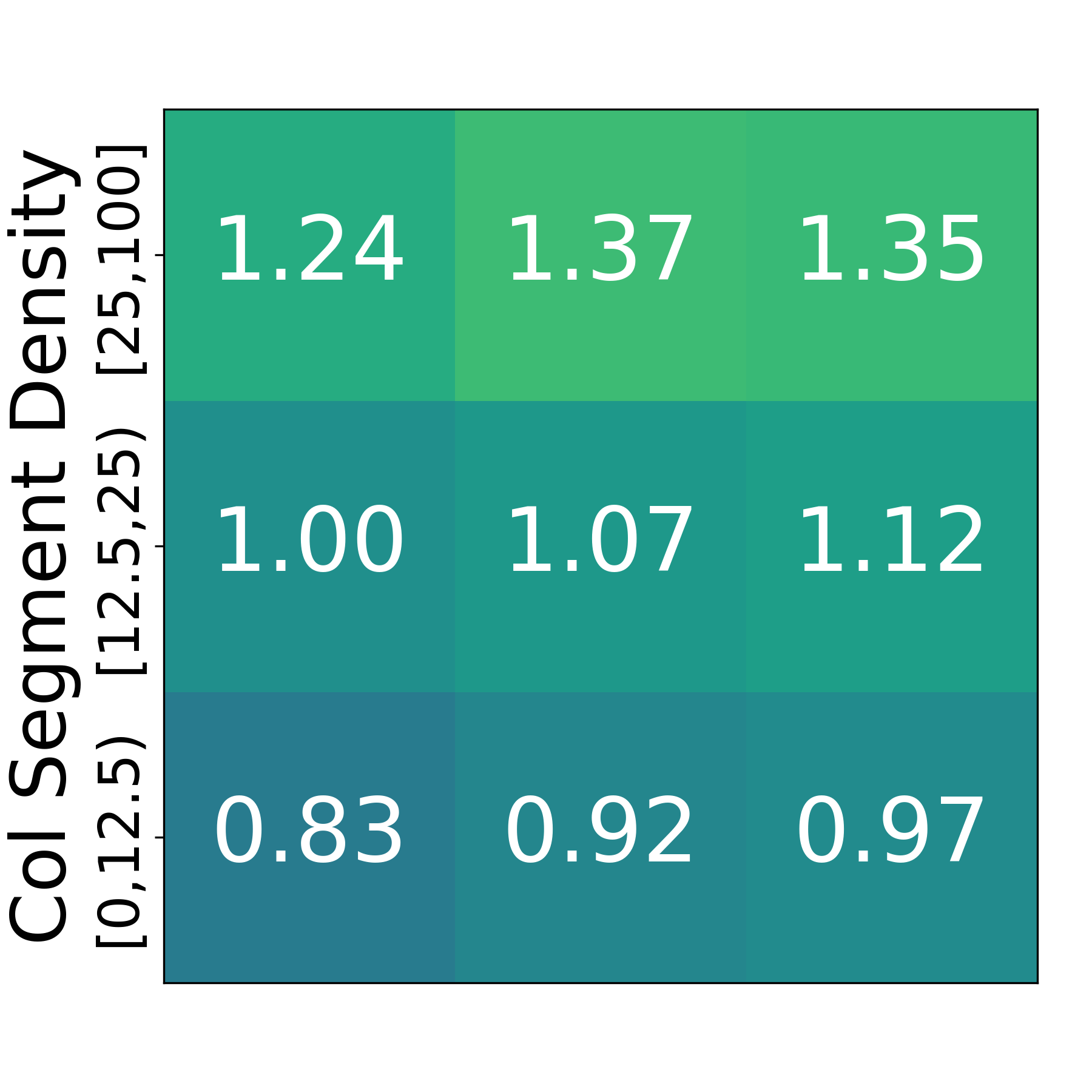}
  \end{subfigure}
  
  \begin{subfigure}[b]{0.49\columnwidth} 
    \includegraphics[width=\linewidth]{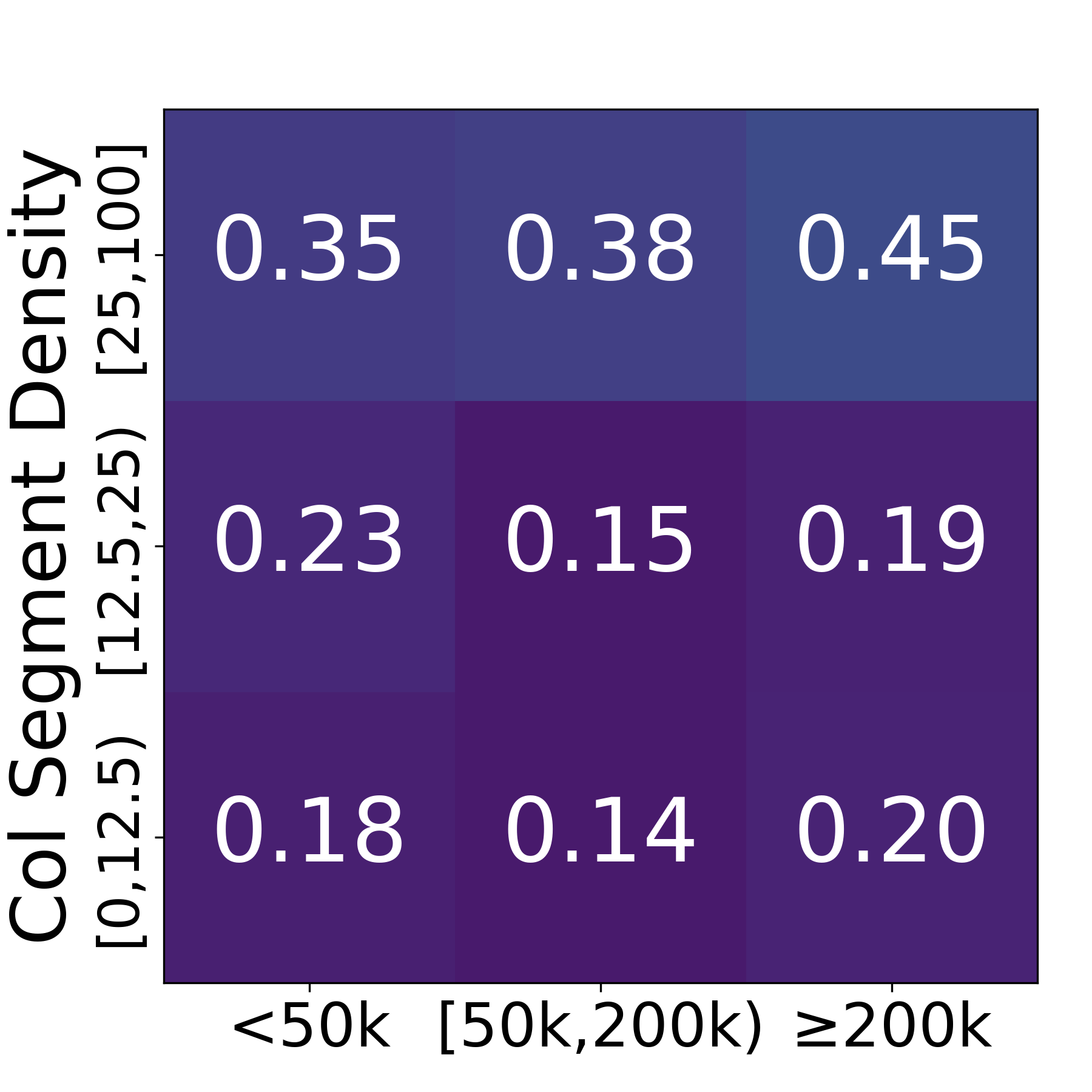}
  \end{subfigure}
  \hfill 
  \begin{subfigure}[b]{0.49\columnwidth} 
    \includegraphics[width=\linewidth]{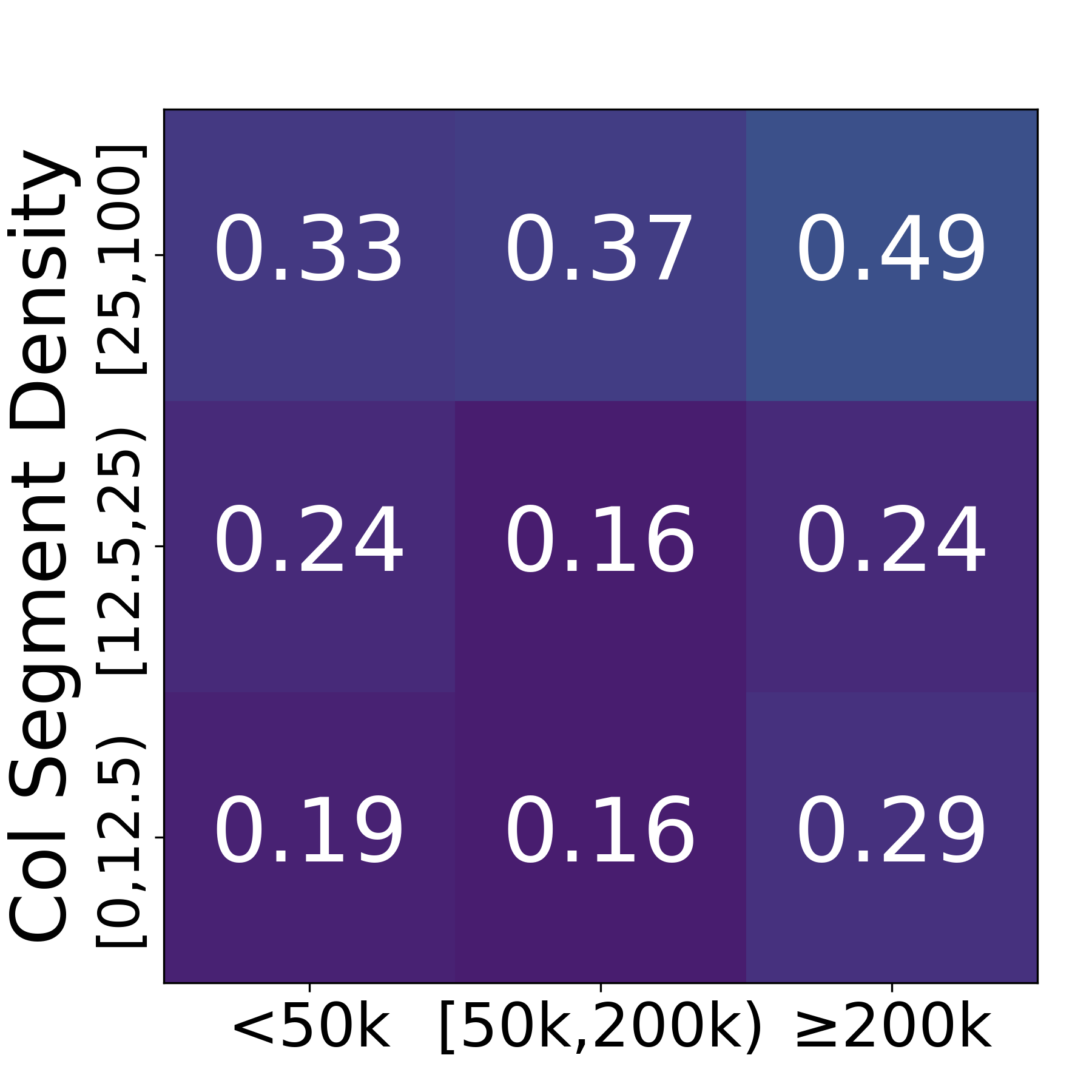}
  \end{subfigure}

  \caption{Speedup heatmap of cuTeSpMM (upper row) and TCGNN (lower row) over Best-SC. Left (Ampere A100) Right (RTX-4090)}
  \label{fig:heatmap}
\end{figure}

\noindent {\bf Hardware:} We present our experimental evaluation on two Nvidia GPUs: an Ampere A100 with 80GB global memory, 108 SMs, and a base clock rate of 765 MHz; and an RTX 4090 with 24GB of global memory, 128 SMs and a base clock rate of 2.2GHz. 

    \noindent {\bf Datasets:} We used all sparse matrices from the SuiteSparse matrix collection \cite{davis2011university,suitesparse} that have more than 10,000 rows/columns. We excluded smaller matrices since GPU acceleration of sparse matrix operations is primarily of interest for sufficiently large matrices, where the overheads of data transfer between host CPU and GPU are much less significant compared to the time for performing the matrix operation. 

\noindent {\bf Compared SpMM Implementations:} In our experimental evaluation, we compare against TC-GNN and multiple SpMM implementations that use scalar cores:
Nvidia cuSparse library \cite{cusparse} with two data representations (CSR and COO), Sputnik \cite{gale2020sparse}, and GE-SpMM \cite{huang2020ge}- a further optimized version with six variants has been incorporated into the dgSPARSE library \cite{dgsparse2021} by the authors (we executed all six SpMM versions in dgSPARSE and chose the fastest one for comparison).
We denote as {\em Best-SC} the performance of the fastest of all of the above scalar-core SpMMs.

\subsection{Performance Measurement Methodology}
We use Nvidia's Nsight-Compute profiling tool \cite{nsight-compute} to accurately collect the kernel execution time for all the methods (e.g., cuSparse CSR, cuSparse COO, Sputnik, Ge-SpMM, TCGNN, and cuTeSpMM). TC-GNN SpMM runs showed two kernels in the Nsight output: one named \texttt{vectorized\_elementwise\_kernel} and another one called \texttt{spmm\_forward\_cuda\_kernel}; we only use the reported  time for \texttt{spmm\_forward\_cuda\_kernel}. For other methods, such as cuSparse-CSR, we aggregated the time for all reported kernels by Nsight-Compute.

\subsection{Pre-processing Overheads} The TCU-based SpMM schemes (cuTeSpMM, 
TC-GNN) require a relatively large pre-processing overhead compared to the time for one SpMM. However, the typical use-cases for SpMM require hundreds to thousands of invocations of SpMM with the same sparse matrix. For example, with GNN training, hundreds of epochs are often needed, and further a number of training runs with various combinations of hyper-parameters like learning rate, dropout etc. are needed. As another example from scientific computing, the LOBPCG \cite{labini2022blocking,anzt2017accelerating} method for iterative solution of a sparse linear system of equations uses SpMM for each iteration until convergence, and the LOBPCG solver can be invoked hundreds to thousands of times with outer loops representing linear iterations to solve a nonlinear system and time stepping through a simulation.

The pre-processing overhead (on the host CPU) for cuTeSpMM is currently around two orders of magnitude higher than the GPU SpMM time for N=128. This overhead is generally lower than that of TC-GNN and also considerably lower than the time to read in the sparse matrix from disk.
\begin{table}[ht]
\centering
\begin{minipage}{0.48\linewidth}
    \centering
    \begin{tabular}{|c|c|}
        \hline
        Synergy & Range \\
        \hline
        Low & $[0\%,12.5\%)$ \\
        Medium & $[12.5\%,25\%)$ \\
        High & $[25\%,100\%]$ \\
        \hline
    \end{tabular}
    \caption{Synergy ranges}
    \label{table:syn_def}
\end{minipage}\hfill
\begin{minipage}{0.48\linewidth}
    \centering
    \begin{tabular}{|l|r|}
        \hline
        \textbf{Synergy} & \textbf{\# of Matrices} \\
        \hline
        Low & 666 \\
        Medium & 198 \\
        High & 235 \\
        \hline
        \textbf{Total} & 1099 \\
        \hline
    \end{tabular}
    \caption{Number of matrices in each synergy category}
    \label{table:synergy_matrices}
\end{minipage}
\end{table}

\subsection{Sparse Matrix Synergy} 
With cuTeSpMM, each active {\em brick} in the HRPB representation has a brick density of nonzero elements that can range from $\frac{4}{64}$ to 1 (each column is guaranteed to have at least one non-zero). Since each brick has four columns, each column in a brick can have a density ranging from $\frac{1}{16}$ to 1. Equivalently, the average number of non-zeros in a column in an active brick can range from 1 to 16. In reporting performance, we classify sparse matrices into three categories with respect to the expected {\em synergy} for TCU-based SpMM:
\begin{compactitem}
    \item {\bf Low Synergy:} These are sparse matrices where the average fraction of nonzeros per column in a brick ($\alpha$) is less than 12.5\% (2/16). These matrices have a low $OI$, with each element of $B$ getting on average no more than one {\em reuse} per shared-memory load instruction. 
    \item {\bf Medium Synergy:} These are sparse matrices with an average active-column non-zero ratio of 12.5\% to 25\% (4/16). For these matrices, $OI_{shared}$ ($512 \times \alpha$) is expected to range between 32 and 64.  
     \item {\bf High Synergy:} Sparse matrices with an average active-column non-zero ratio greater than 25\%. We expect such matrices to benefit significantly from TCUs, with $OI_{shared}$ expected to exceed 64. 
\end{compactitem}

Table~\ref{table:syn_def} summarizes the definition of the three synergy levels, while table~\ref{table:synergy_matrices} shows the distribution of the tested Suitesparse matrices in terms of these three degrees of synergy for TCU-based SpMM execution.

\subsection{Observations on Performance}
Fig.~\ref{fig:combined_boxplots} shows box-plots showing the distribution of performance (TFLOPs) on the A100 (left) and 4090 (right), with the matrices split along the three synergy classes: High/Medium/Low.  We make a few summary observations:
\begin{itemize}
    \item {\bf Consistently high performance compared to TC-GNN:} Across all three synergy categories and for all dense matrix widths (32/128/512), cuTeSpMM (leftmost box in the groups of three) is consistently better with respect to minimum, 25th, 50th, and 75th percentiles and maximum achieved performance. In virtually all cases, the TFLOP performance at the 25th percentile for cuTeSpMM is higher than the 75th percentile performance of TC-GNN.
    \item {\bf Significantly higher performance over Best-SC for High-Synergy matrices}: The top set of box plots show that the performance of cuTeSpMM is consistently and significantly higher (especially for maximum, but also for 25th, 50th, and 75th percentiles) than Best-SC.
    \item {\bf Very good performance relative to Best-SC even for Medium/Low synergy matrices:} For medium-nad low- synergy matrices, for most cases cuTeSpMM has better median performance (middle red bars in box plots) and also has better 25th and 75th percentile performance than Best-SC (except for N=32). 

\end{itemize}

Fig.~\ref{fig:heatmap} shows the relative perform of cuTeSpMM as compared to Best-SC (left) and TC-GNN as compared to Best-SC (right). Matrices with fewer rows are more likely to have load imbalance, and matrices with low synergy are expected to not benefit as much from TC-GNN. The cuTeSpMM strongly exhibit both of these trends for both the A100 and RTX 4090. TC-GNN does exhibits similar relative trends (high synergy general has a better speedup than low synergy, and matrices with many rows general have a higher speedup). However, TC-GNN has a speedup of less than 0.5 for all groups for both GPUs. Therefore, the expected TCU performance trends are seen for both TC-GNN and cuTeSpMM even though TC-GNN does not perform well for these experiments.

\section{Related Work}
\label{sec:related}

\subsection{Sparse Tensor Core}
\label{sparse-tc}

The NVIDIA Ampere Architecture GPU has added support for sparse tensor cores\cite{sparseTc}. These cores are designed to accelerate the inference of machine learning models where the weight matrix has been pruned to a specific 2:4 sparse pattern. 
This pattern requires two out of every four values to be zero, making it unsuitable for all sparse matrices. In this work, we target general sparse matrices, especially highly sparse matrices with well over 99\% sparsity.

\subsection{TC-GNN}
\label{TC-GNN}
In their work \cite{wang2023tc}, Wang et al. introduced the TC-GNN SpMM kernel to expedite the training process of GNNs.
They focused primarily on matrices arising in GNN use-cases. In contrast, we address sparse matrices from across various domains. We have compared with TC-GNN in our work and observed significant speedups across all matrix categories. 

\subsection{FLASH LLM}
\label{flash_llm}
Recently, Xia et al. introduced $FLASH\_LLM$ SpMM in their paper \cite{xia2023flash}, leveraging Tensor Cores to accelerate the SpMM operation within the machine learning domain. Notably, the matrices they evaluated were self-generated random matrices with densities ranging from 10\% to 30\%, which are considerably denser than the matrices typically encountered in GNN/scientific computing, our primary focus. Additionally, they employed a different data type, FP16, while we utilize FP32 (TF32). The matrices we target tend to have densities below 1\%, presenting further challenges such as low computation efficiency (low brick density) and load imbalance (non-zeros are not uniformly distributed), which must be addressed to achieve optimal performance. We were unable to directly compare performance with their implementation due to the difference in data precision targeted. Further, the domain they target features sparse matrices with very much lower sparsity, typically between 60\% to at most 95\%.


%
\section{Conclusion}
\label{sec:conclusion}
In this paper, we have developed a new sparse matrix data representation and an efficient GPU kernel to make effective use of GPU tensor cores for SpMM. We have demonstrated significant performance improvements over existing TCU-based SpMM implementations as well as significant performance improvements even over scalar-core based SpMM implementations for sparse matrices that can be characterized as having high TCU-synergy.

\begin{acks}
This work was supported in part by the U.S. National Science Foundation through award 2009007.
\end{acks}

\label{sec:plots}

%
\clearpage

\end{document}